\documentstyle[psfig]{mn} 
\title{The intermediate-age open cluster 
 NGC~2158\thanks{Based on observations  
carried out at Mt Ekar, Asiago, Italy. All the photometry is available 
at WEBDA database: http://obswww.unige.ch/webda/navigation.html}} 
\author[Carraro et al.]        
{Giovanni Carraro$^1$, 
L\'eo Girardi$^{1,2}$ and  
Paola Marigo$^1$\thanks{email: 
giovanni.carraro@unipd.it (GC); lgirardi@pd.astro.it (LG); 
marigo@pd.astro.it (PM)}\\ 
$^1$ Dipartimento di Astronomia, Universit\`a di Padova, Vicolo 
dell'Osservatorio 2, I-35122 Padova, Italy \\ 
$^2$ Osservatorio Astronomico di Trieste, Via G.B.\ Tiepolo 11,  
I-34131 Trieste, Italy \\ 
} 
 
\date{\it Submitted: October 2001} 
\pubyear{2002} 
\begin{document} 
\maketitle 
\title{The open cluster NGC~2158} 
 
\begin{abstract} 
We report on $UBVRI$ CCD photometry of two overlapping fields  
in the region of  
the intermediate-age open cluster NGC~2158 down to $V=21$. 
By analyzing Colour-Colour (CC) and Colour-Magnitude Diagrams (CMD) we  
infer a reddening $E_{B-V}= 
0.55\pm0.10$, a distance of $3600 \pm 400$ pc, and an age 
of about 2 Gyr. Synthetic CMDs performed with these parameters 
(but fixing $E_{B-V}=0.60$ and $[{\rm Fe/H}]=-0.60$),  
and including binaries, field contamination, and photometric errors, 
allow a good description of the observed CMD. 
The elongated shape of the clump of red giants in the CMD 
is interpreted as resulting from a differential reddening of about 
$\Delta E_{B-V}=0.06$ across the cluster, in the direction 
perpendicular to the Galactic plane.   
NGC~2158 turns out to be an intermediate-age open cluster 
with an anomalously low  metal  content.  
The combination of these 
parameters together with the analysis of the cluster orbit,  
suggests that the cluster belongs to the old thin disk population. 
\end{abstract} 
 

\section{Introduction} 
NGC~2158 (OCL~468, Lund~206, Melotte~40)  
is a rich northern open cluster of intermediate age, 
located low in the galactic plane toward 
the anti-center direction ($\alpha=06^{\rm h}~07^{\rm m}.5$,  
$\delta=+24^{\circ} 
06^{\prime}$, $\ell=186^{\circ}.64$, $b=+1^{\circ}.80$, J2000.),  
close to M~35. 
It is classified as a II3r open cluster by Trumpler 
(1930), and has a diameter of about $5^{\prime}$, according to Lyng\aa\ (1987). 
It is quite an interesting object due to its shape, for which in the  
past it was considered a possible globular cluster, also presenting an  
unusual combination of age and metallicity. In fact it is an  
intermediate-age open cluster, but rather metal poor. 
It is a crucial object in determining the  
Galactic disk abundance gradient and the abundance spread  
at time and place in the disk. 
 
The cluster is rather populous,  
and therefore it is an ideal candidate 
to be compared with theoretical models of intermediate-low mass stars 
(Carraro \& Chiosi 1994a, Carraro et al. 1999). 
Since in the past no detailed studies have been pursued 
with this aim, we decided to undertake a multicolor  
CCD study of the cluster, which is presented 
in the present paper. 
Moreover this paper is the third of a series dedicated at improving 
the photometry of northern intermediate-age open clusters at Asiago 
Observatory. We already reported elsewhere on NGC~1245 (Carraro \& Patat 1994) 
and on NGC~7762 (Patat \& Carraro 1995). 
 
The plan of the paper is as follows. In Sect.~2 we summarize 
the previous studies on NGC~2158, while Sect.~3 is dedicated  
to present the observation and reduction strategies. 
The analysis of the CMD is performed in Sect.~4, whereas 
Sect.~5 deals with the determination of cluster reddening, 
distance and age. Sect.~6 illustrates NGC~2158 kinematics.  
Finally, Sect.~7 summarizes our findings. 
 
\begin{figure*} 
\centerline{\psfig{file=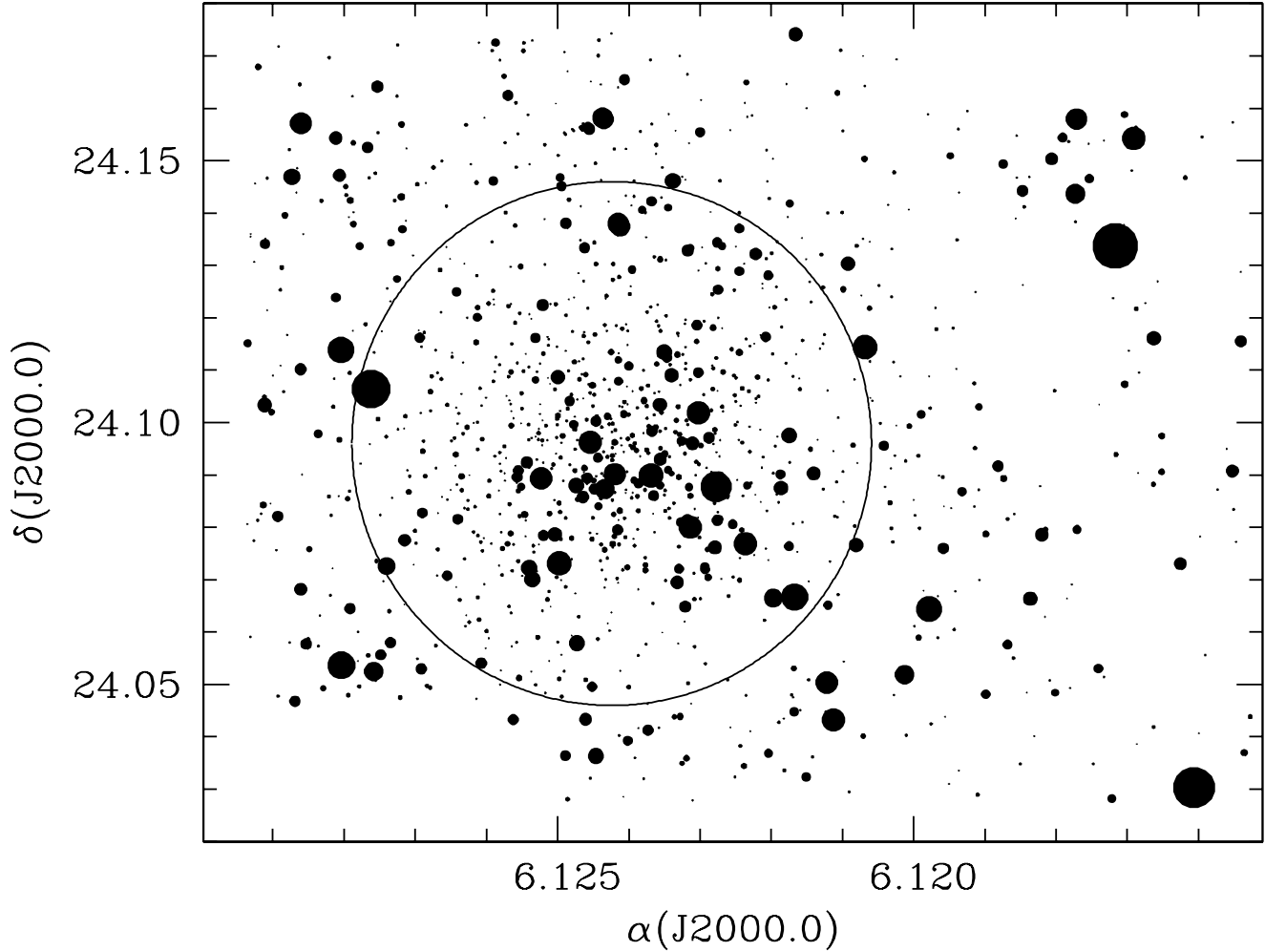,width=\textwidth}} 
\caption{A $V$ map of the observed field from the  
photometry of one of the deep  
$V$ frames; North is up and East is to the left; the field is  
$9 \times 11 \,{\rm arcmin}^{2}$. 
The circle confines the stars within $3^{\prime}$ from the cluster center. 
The size of each star is inversely proportional to its magnitude.} 
\label{mappa} 
\end{figure*} 
 
\section{Previous investigations} 
NGC~2158 has been studied several times in the past. 
The first investigation was carried out by Arp \& Cuffey (1962), 
who obtained photographic $BV$ photometry for about 900 
stars down to $V=18.5$. Photographic photometry was also obtained by 
Karchenko et al (1997) for more than 2000 stars down to the same 
limiting magnitude together with proper motions. 
 
CCD photometry in $BV$ passbands was provided by Christian et al. (1985) 
and Piersimoni et al. (1993).  Both these studies  
reach deeper magnitude limits. Anyhow, the former study basically provides 
only a selection of MS unevolved stars, whereas the latter one 
presents a nice CMD, but the analysis of the data appears very 
preliminary. 
 
There is some disagreement in  
the literature about the value of NGC~2158 fundamental 
parameters, specially with respect to the cluster 
age. 
Estimates of cluster metallicities have been obtained by several authors, 
and, although different, they all point to a sub-solar 
metal content ($[{\rm Fe/H}]=-0.60$,  Geisler 1987, Lyng\aa\ 1987). 
Finally, the kinematics of NGC~2158 has been studied by measuring  
spectra of giant stars (Scott et al. 1995; Minniti 1995) to provide 
radial velocities. It turns out that the mean cluster radial velocity  
is in the range $15-30$ km/s (Scott et al. 1995).  
 
\begin{table} 
\tabcolsep 0.30truecm 
\caption{Journal of observations of NGC~2158 (January 6-7 , 2000).} 
\begin{tabular}{cccc} 
\hline 
\multicolumn{1}{c}{Field}    & 
\multicolumn{1}{c}{Filter}    & 
\multicolumn{1}{c}{Time integration}& 
\multicolumn{1}{c}{Seeing}         \\ 
      &        & (sec)     & ($\prime\prime$)\\ 
  
\hline 
 $\#1$   &   &      &      \\ 
         & $U$ &  240 &  1.2 \\ 
         & $B$ &  300 &  1.3 \\ 
         & $V$ &  120 &  1.3 \\ 
         & $R$ &   60 &  1.5 \\ 
         & $I$ &  120 &  1.3 \\ 
 $\#2$   &   &      &      \\ 
         & $U$ &  240 &  1.2 \\ 
         & $B$ &  300 &  1.1 \\ 
         & $V$ &  120 &  1.3 \\ 
         & $R$ &   60 &  1.5 \\ 
         & $I$ &  120 &  1.3 \\ 
\hline 
\end{tabular} 
\end{table}

\begin{figure}  
\centerline{\psfig{file=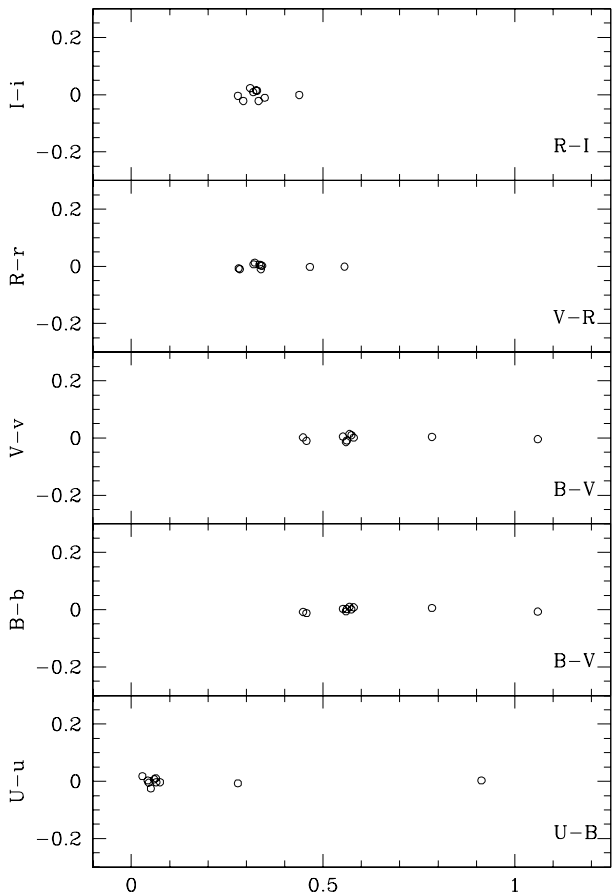,width=\columnwidth}} 
\caption{Differences between standard magnitudes and those obtained  
from our Eq.~\protect\ref{eq_calib}, for our standard stars and as a  
function of colour.} 
\label{fig_standards} 
\end{figure} 
 
\begin{figure}  
\centerline{\psfig{file=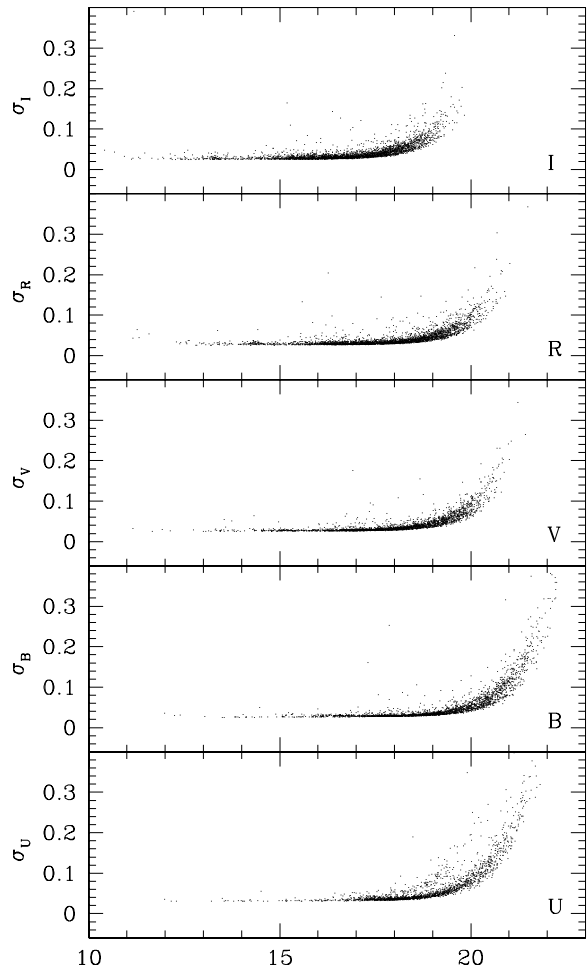,width=\columnwidth}} 
\caption{Photometric errors as a function of magnitude, for our NGC~2158 
observations.} 
\label{fig_errors} 
\end{figure} 
 
\section{Observations and Data Reduction} 
 
Observations were carried out with the AFOSC camera at the 
1.82~m telescope of Cima Ekar, in the nights of January 6 and 7, 
2000. AFOSC samples a $8^\prime.14\times8^\prime.14$ field in a  
$1K\times 1K$ thinned CCD. The typical seeing was between 1.0  
and 1.5 arcsec.  
 
For NGC~2158, typical exposure times were of 240~s in $U$,  
300~s in $B$, and 60--120~s in $VRI$. Several images  
were taken, either centered on the cluster core, or shifted by about 
$4^\prime$ in order to better sample the neighboring field  
(see Fig.~\ref{mappa}). However, 
only the images with the best seeing were used.   
We also observed a set of standard stars in M~67 (Schild 1983;  
and Porter, unpublished). 
 
The data has been reduced by using the IRAF\footnote{IRAF  
is distributed by the National Optical Astronomy Observatories, 
which are operated by the Association of Universities for Research 
in Astronomy, Inc., under cooperative agreement with the National 
Science Foundation.} packages CCDRED, DAOPHOT, and PHOTCAL. 
The calibration equations obtained (see Fig.~\ref{fig_standards}) are: 
	\begin{eqnarray}  
\nonumber 
u \! &=& \! U + 4.080\pm0.005 + (0.010\pm0.015)(U\!-\!B) + 0.55\,X \\  
\nonumber 
b \! &=& \! B + 1.645\pm0.010 + (0.039\pm0.015)(B\!-\!V) + 0.30\,X \\  
\nonumber 
v \! &=& \! V + 1.067\pm0.011 - (0.056\pm0.018)(B\!-\!V) + 0.18\,X \\  
\nonumber  
r \! &=& \! R + 1.109\pm0.012 - (0.075\pm0.032)(V\!-\!R) + 0.13\,X \\  
\nonumber 
i \! &=& \! I + 1.989\pm0.048 + (0.118\pm0.145)(R\!-\!I) + 0.08\,X \\ 
	\label{eq_calib} 
	\end{eqnarray} 
where $UBVRI$ are standard magnitudes, $ubvri$ are the instrumental  
ones, and $X$ is the airmass. For the extinction coefficients, 
we assumed the typical values for the Asiago Observatory. 
Figure~\ref{fig_standards} shows the residuals of the above  
calibration equations as a function of colour for all our standard  
stars.  
 
\begin{figure*} 
\centerline{\psfig{file=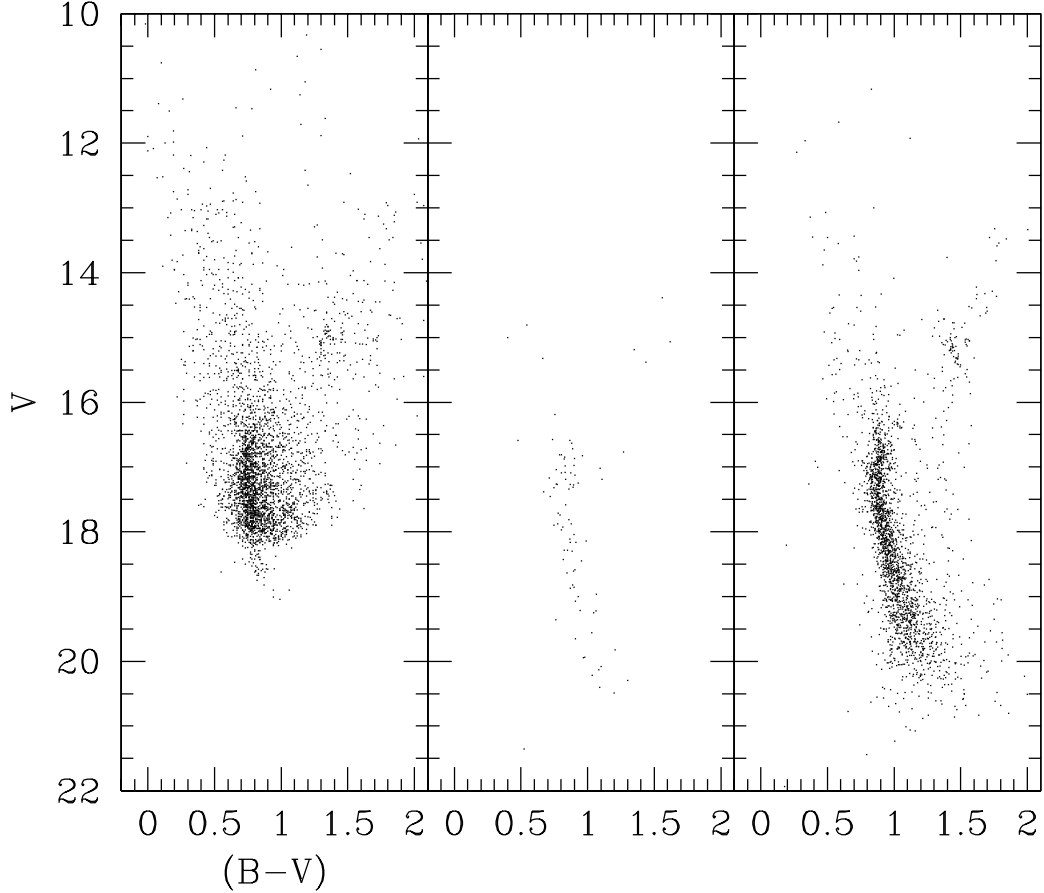,width=0.8\textwidth}} 
\caption{$BV$ CMDs of NGC~2158. The left panel presents the Arp \& Cuffey 
(1962) photometry, the central panel the Christian et al. (1985) photometry,  
whereas the right panel shows our photometry.} 
\end{figure*} 
 
Finally, Fig.~\ref{fig_errors} presents the run of photometric errors 
as a function of magnitude. These errors take into account fitting errors 
from DAOPHOT and calibration errors, and have been computed 
following Patat \& Carraro (2001). 
It can be noticed that stars brighter than  
about 20 in $V$, $R$, and $I$, 21 in $B$, and $U$, have  
photometric errors lower than 0.1~mag. The final photometric data is  
available in electronic form at the  
WEBDA\footnote{http://obswww.unige.ch/webda/navigation.html} site.

\section{The Colour-Magnitude Diagrams} 
A comparison of our photometry with past analyses is shown in Fig.~4, 
from which it is evident that the present study supersedes the 
previous ones. In fact, we reach $V=21$, and are able to cover 
all the relevant regions of the CMD. Instead, 
Arp \& Cuffey (1958) photometry extends only for a couple 
of magnitudes below the turn-off point (TO) , whereas the photometry 
of Christian et al. (1985) does not cover the evolved region of the CMD. 
 
To better identify the TO location and the Red Giant (RG) 
clump, in Fig.~5 we plot the CMDs obtained by considering stars 
located in different cluster regions. 
In details,  
left panel presents the CMD obtained by including all the measured stars, 
central panel considers the stars within a circle of radius $3^{\prime}$,  
whereas the right panel presents only the stars located inside a circle of  
radius $1.5^{\prime}$. 
The radius adopted in the central panel  
is compatible with the available estimate of the  
cluster diameter, which is about $5^{\prime}$, so that we 
are likely considering most of the cluster members. 
By inspecting this CMD, we find that the TO is located at 
$V \approx 16.0$, $(B-V) \approx 1.0$, whereas a prominent clump 
of He burning stars is  visible at $V \approx 15.0$, $(B-V) \approx 1.5$. 
The diagonal structure of this latter is probably due to differential 
reddening effects, which we are going to discuss in Sect.~5.3. 
The MS extends for 5 magnitudes, getting wider at increasing 
magnitudes: this is compatible with the trend of photometric 
errors (see Fig.~3) and the probable presence of a significant  
population of binary stars. 
The global CMD morphology resembles that of NGC~7789 (Vallenari et al. 2000) 
and  NGC~2141 (Carraro et al. 2001), two  
well studied rich intermediate-age open clusters.

\begin{figure*} 
\centerline{\psfig{file=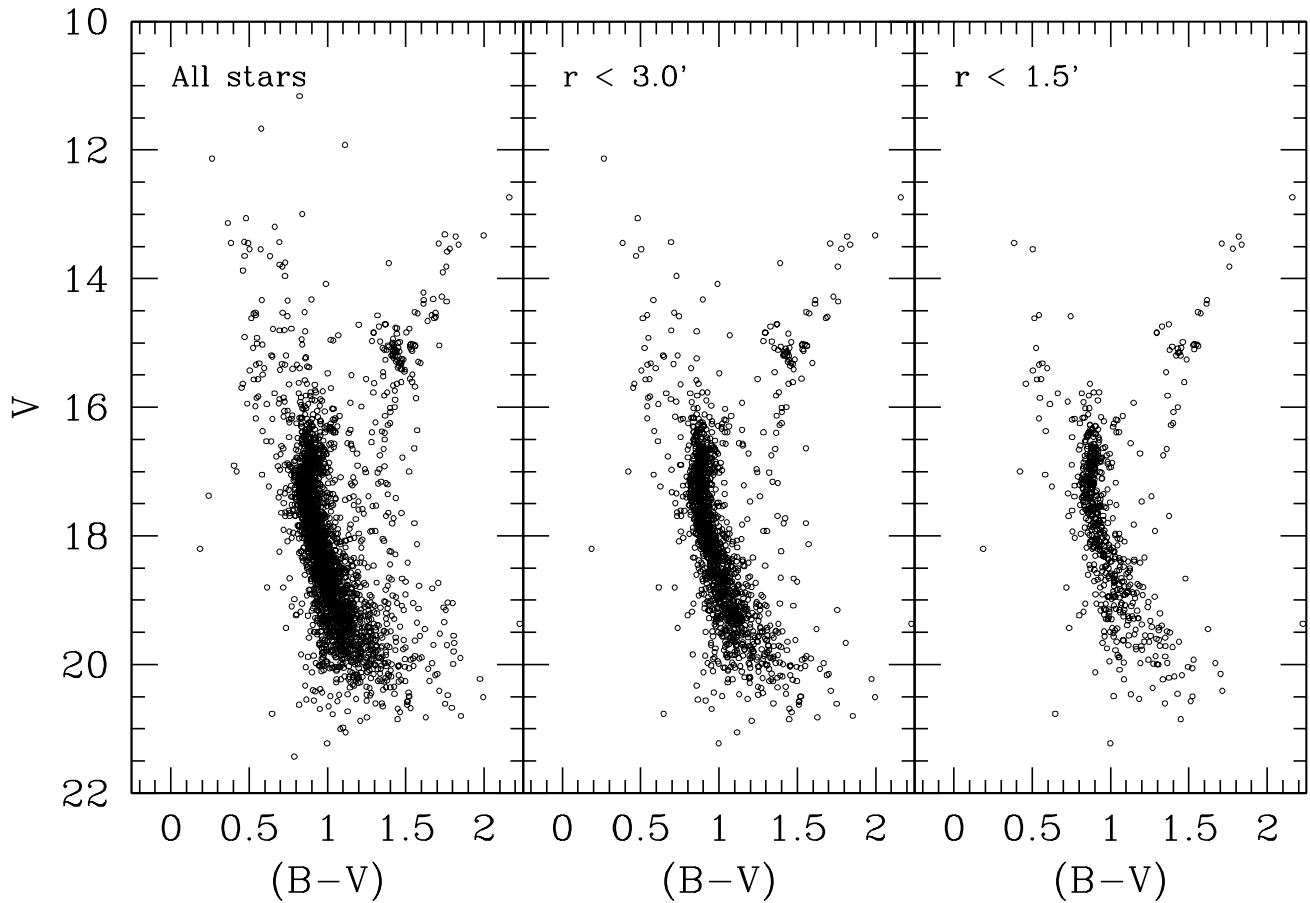,width=\textwidth}} 
\caption{$BV$ CMDs of NGC~2158.  
The left panel presents the CMD obtained by including all the measured stars, 
the central panel considers the stars within a radius of $3^\prime$, whereas 
the right panel shows only the stars located inside a radius of $1.5^\prime$.} 
\end{figure*} 
 
\begin{table*}  
\tabcolsep 0.7truecm 
\caption {NGC~2158 fundamental parameters taken from the literature.} 
\begin{tabular}{ccccc}  
\hline 
&Arp \& Cuffey & Christian et al. & Kharcenko et al. & Piersimoni et al. \\ 
\hline 
$E(B-V)$   &  0.43  &  0.55 &  0.35 &  0.55 \\ 
$(m-M)$    & 14.74  & 14.40 & 13.90 & 15.10 \\ 
distance (pc) & 4700   & 3500  & 3700  & 4700\\ 
Age (Gyr)  & 0.8    & 1.5   & 3.0   & 1.2 \\  
\hline 
\end{tabular} 
\end{table*} 
 
\section{Cluster fundamental parameters} 
There fundamental parameters of NGC~2158 are still controversial in the  
literature (see Table~2). 
The cluster age estimates range from 0.8 to 3.0 Gyr, 
the distance from 3500 to 4700~pc and the reddening $E_{B-V}$ from 0.35 
to 0.55. 
In the next sections we are going to derive update estimates for 
NGC~2158 basic parameters. 
 
\subsection{Reddening} 
\label{sec_reddening} 
In order to obtain an estimate of the cluster mean reddening, we 
analyse the distribution of the stars with $V < 17$ in the $(B-I)$  
vs. $(B-V)$ plane, which is shown in Fig.~6. 
 
The linear fit to the main sequence in the $(B-I)$  
vs. $(B-V)$ plane, 
\begin{equation} 
(B-I) = Q + 2.25 \times (B-V) 
\end{equation} 
can be expressed in terms of $E_{B-V}$, for the $R_V=3.1$ 
extinction law, as 
\begin{equation} 
E_{B-V} = \frac{Q-0.014}{0.159} \,\,\,\,  , 
\label{mu2} 
\end{equation} 
following the method proposed by  Munari \& Carraro (1996a,b). 
This method provides a rough estimate of the mean reddening and, 
as amply discussed in Munari \& Carraro (1996a), can be used 
only for certain colour ranges. In particular Eq.~(\ref{mu2}) 
holds over the range $-0.23 \leq (B-V)_0 \leq +1.30$. 
MS stars have been selected by considering all the stars within 
$3^{\prime}$ from the cluster center and having $17 \leq V \leq 21$ and  
$0.75 \leq (B-V) \leq 1.25$. 
A least-squares  fit through all these stars 
gives $Q=0.097$, 
which, inserted in Eq.~(\ref{mu2}), provides  $E_{B-V}=0.56\pm0.17$. 
The uncertainty is rather large, and is due to the scatter 
of the stars in this plane, which indicates the presence 
of stars with different reddening, presumably a mixture 
of stars belonging to the cluster and to the field. 
 
Another indication of the cluster mean reddening can be derived 
from the Colour-Colour diagram $(U-B)$ vs. $(B-V)$, shown 
in Fig.~7. 
Here we consider again all the stars located within $3^\prime$  
from the cluster center having $17 \leq V \leq 21$ and  
$0.75 \leq (B-V) \leq 1.25$, to alleviate the contamination effect. 
The solid line is an empirical Zero-Age 
Main Sequence (ZAMS) taken from Schmidt-Kaler (1982), 
whereas the dashed line is the same ZAMS, but shifted by 
$E_{B-V}=0.55$. The ratio $E_{U-B}/E_{B-V}=0.72$ has  
been adopted. This shift, together with the dispersion of the  
data around the shifted ZAMS, provides the reddening value 
of $E_{B-V}=0.55\pm0.10$. 
 
\subsection{Distance and age} 
\label{sec_distanceage} 
 
As already mentioned, there is still a considerable  
dispersion in the literature  
among different estimates of NGC~2158 distance and age. 
We have derived new estimates for these parameters as follows. 
 
First, from the Girardi et al.\ (2000a) database we generate  
theoretical isochrones of metallicity $Z=0.0048$, that corresponds  
to the observed value of $[{\rm Fe/H}]=-0.60$. Figure~\ref{fig_isocdata} 
shows the isochrones with ages between 1.58 to 
2.51 Gyr, which defines the age interval compatible  
with the observed magnitude difference between the red clump and 
the turn-off region. The isochrones were shifted  
in apparent magnitude and colour, until the locus of  
core-helium burning stars coincided with the observed  
mean position of the clump.  
The results, shown in Fig.~\ref{fig_isocdata}, 
imply a true distance modulus of $(m-M)_0=12.8$~mag  
(3630~pc), and a colour excess of $E_{B-V}=0.60$ for NGC~2158.  
This value is compatible with the one obtained in the previous 
Sect.~\ref{sec_reddening}.

\begin{figure} 
\centerline{\psfig{file=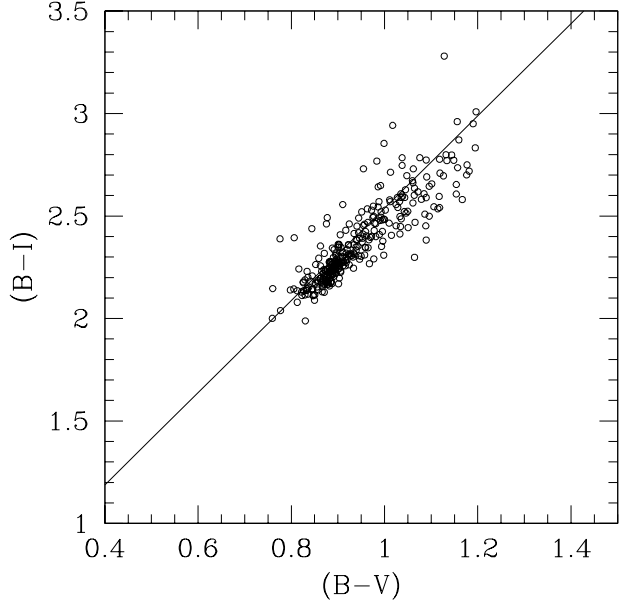,width=\columnwidth}} 
\caption{NGC~2158 MS stars within $3^\prime$ 
in the $(B-V)$ vs. $(B-I)$ plane.} 
\end{figure} 
 
\begin{figure} 
\centerline{\psfig{file=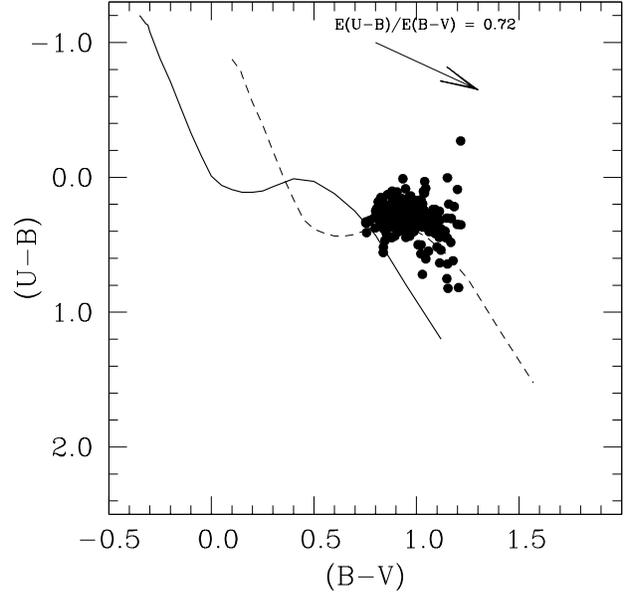,width=\columnwidth}} 
\caption{NGC~2158 stars within $3^\prime$ in the colour-colour 
diagram. The solid line is an empirical ZAMS taken from 
Schmidt-Kaler (1982), whereas the dashed line is the same ZAMS, 
but shifted by $E_{B-V}=0.55$. The arrow indicates the 
reddening law. } 
\end{figure} 
 
\begin{figure} 
\centerline{\psfig{file=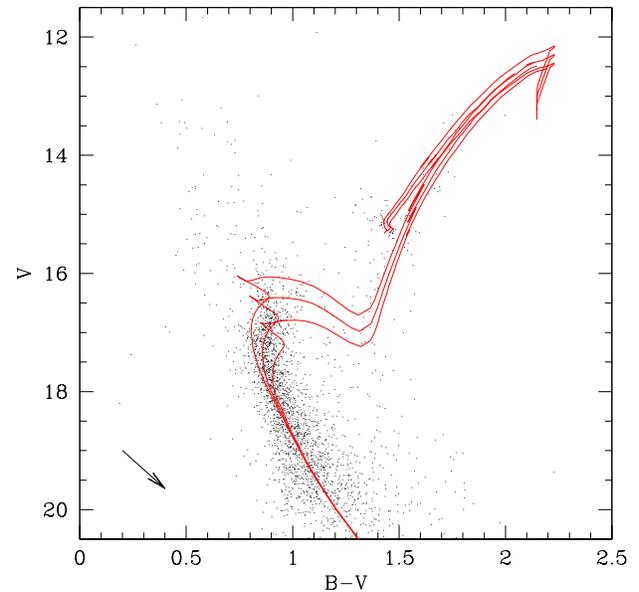,width=\columnwidth}} 
\caption{ NGC~2158 data the $V$ vs.\ $B-V$ diagram (points),  
as compared to Girardi et al.\ (2000a) isochrones of ages 
$1.58\times10^9$, $2.00\times10^9$, and $2.51\times10^9$ yr 
(solid lines), for the metallicity $Z=0.0048$. A distance 
modulus of $(m-M)_0=12.8$, and a colour excess of $E_{B-V}=0.60$, 
have been adopted. The direction of the reddening vector is indicated 
by the arrow at the bottom left. 
} 
\label{fig_isocdata} 
\end{figure} 
 
\begin{figure*} 
\centerline{\psfig{file=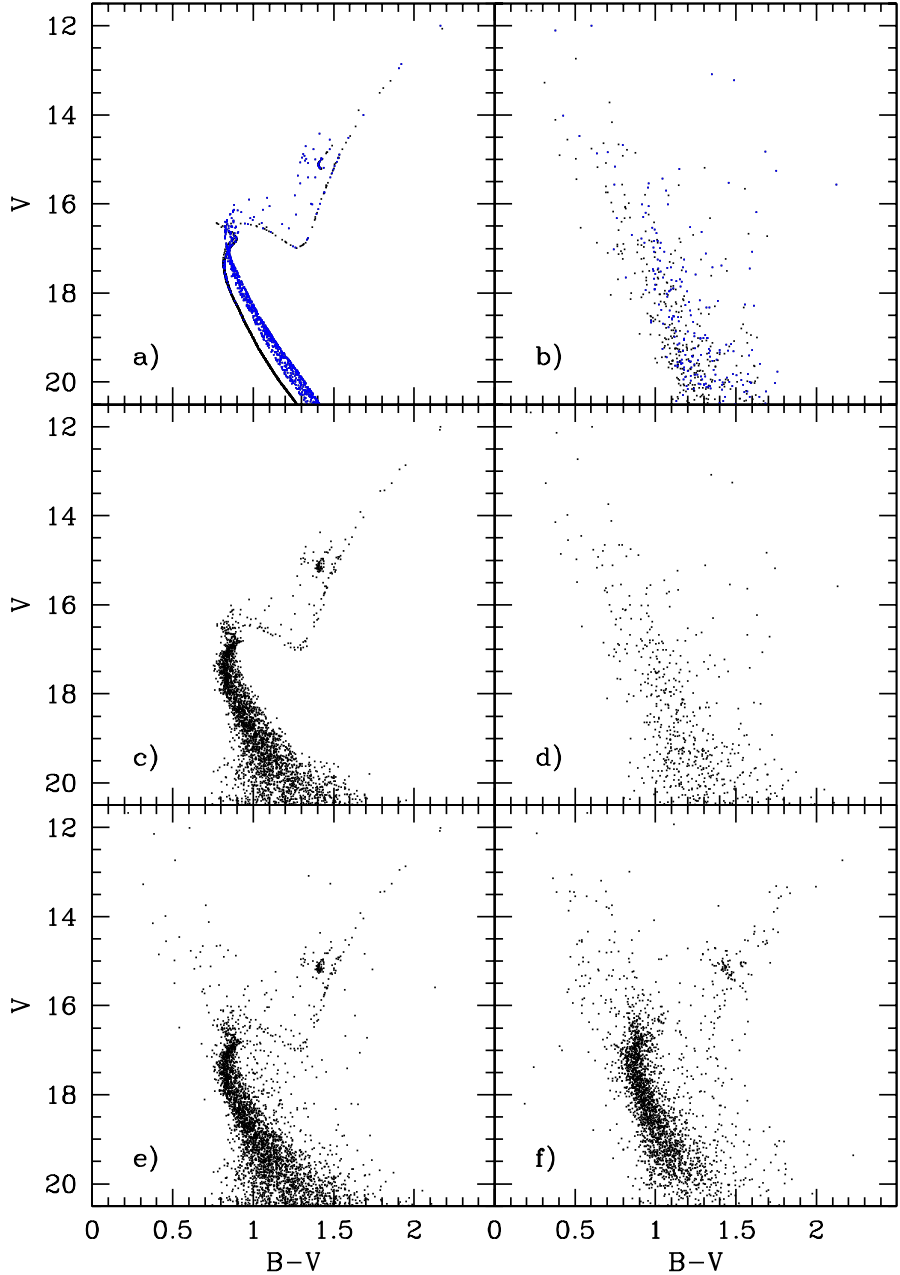,width=0.7\textwidth}} 
\caption{ Simulations of NGC~2158 and its field in the $V$ vs.\ $B-V$ 
diagram.  
{\bf (a)} Simulation of a $2$-Gyr old cluster with $Z=0.0048$ 
and a initial mass of $1.5\times10^4$~$M_\odot$, based on the same  
isochrones, distance modulus and colour excess as in  
Fig.~\protect\ref{fig_isocdata}. We have assumed that 
30 percent of the stars are binaries with mass ratios between 0.7  
and 1.0.  
{\bf (b)} Simulation of a $9 \times 11 {\rm arcmin}^{2}$ field centered at 
galactic coordinates $\ell=186^\circ.64$, $b=+1^\circ.80$, performed with 
Girardi et al.\ (2001) Galactic model. 
Panels {\bf (c)} and {\bf (d)} are the same as (a) and (b), 
respectively, after simulation of photometric errors. 
Panel {\bf (e)} shows the sum of (c) and (d), that can be  
compared to the observational data shown in panel {\bf (f).}  
} 
\label{fig_simuldata} 
\end{figure*} 
 
\begin{figure*} 
\centerline{\psfig{file=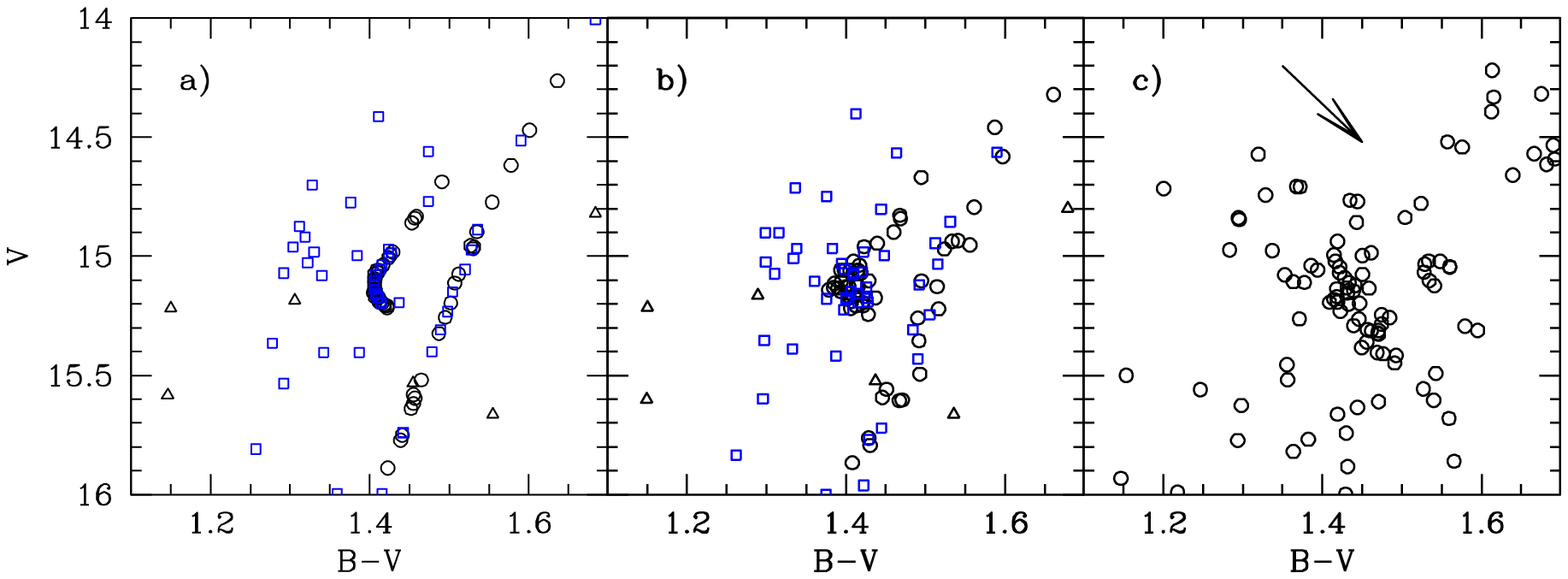,width=\textwidth}} 
\caption{ The same simulations of Fig.~\protect\ref{fig_simuldata}, 
but detailing the region of clump giants. 
{\bf (a)} The simulated data. Circles are single stars in the cluster,  
squares are the binaries, and the few triangles are field stars. 
{\bf (b)} The same as in panel (a), but including the simulation  
of photometric errors. 
{\bf (c)} The observed data for NGC~2158. The arrow shows the reddening 
vector corresponding to $\Delta E_{B-V}=0.1$.  
} 
\label{fig_simulclump} 
\end{figure*} 
 
It should be remarked that these are just first estimates of the  
cluster parameters,  
that we will now try to test further by means of synthetic CMDs. 
Figure~\ref{fig_simuldata} shows the sequence of steps required to 
simulate a CMD aimed to reproduce the NGC~2158 data. 
These steps are: 
	\begin{itemize} 
	\item  
The 2-Gyr old isochrone of $Z=0.0048$ is used to simulate a 
cluster with 100 red clump stars. Assuming a Kroupa (2001) IMF, 
in order to reach this number we need an initial cluster mass of 
about $1.5\times10^4$~$M_\odot$, which is assumed herein-after. 
We have simulated detached binaries, assuming that 30 percent 
of the observed objects are binaries with a mass ratio located 
between 0.7 and 1.0. This prescription is in agreement with 
several estimates for galactic open clusters (Carraro \& Chiosi 19984ab) 
and with the observational data for NGC~1818 and NGC~1866 in  
the LMC (Elson et al.\ 1998; Barmina et al. 2001).  
The result of such simulation is shown in Fig.~\ref{fig_simuldata}(a). 
In this panel we can see that most of the stars -- the 
single ones -- distribute along the very thin sequence defined 
by the theoretical isochrone. Binaries appear as both (i) a sequence of 
objects roughly parallel to the main sequence of single stars, and 
(ii) some more scattered objects in the evolved part of the CMD. 
	\item 
In order to estimate the location of foreground and background  
stars, we use a simple Galaxy model code  
(Girardi et al., in preparation). It includes   
the several Galactic components -- thin and thick disk, halo,  
and an extinction layer -- adopting geometric parameters as  
calibrated by Groenewegen et al.\ (2001). The most relevant component  
in this case is the thin disk, which is modelled by  
exponential density distributions in both vertical and radial 
directions. The radial scale heigth is kept fixed (2.8 kpc),  
whereas the vertical scale heigth $h_z$ increases with the stellar  
age $t$ as 
	\begin{equation} 
h_z = z_0 (1+t/t_0)^\alpha 
	\end{equation} 
with $z_0=95$ pc, $t_0=4.4$ Gyr, $\alpha=1.66$. 
The simulated field has the same  
area ($9 \times 11\, {\rm arcmin}^{2}$) and galactic coordinates  
($\ell=186^\circ.64$, $b=+1^\circ.80$) as the observed one for  
NGC~2158. The results are shown in Fig.~\ref{fig_simuldata}(b).  
It is noteworthy that, in this direction, most of the Galactic  
field stars appear in a sort of diagonal sequence in the CMD,  
that roughly corresponds to the position of NGC~2158 main sequence. 
	\item 
We then simulate the photometric errors as a function of $V$  
magnitude, with typical values derived from our observations  
(see Fig.~\ref{fig_errors}). The results are shown separately for cluster 
and field stars in panels (c) and (d) of Fig.~\ref{fig_simuldata}.  
	\item  
The sum of field and cluster simulations is shown in  
Fig.~\ref{fig_simuldata}(e). This can be compared directly to the 
observed data shown in Fig.~\ref{fig_simuldata}(f). 
	\end{itemize} 
The comparison of these two latter panels indicates that the selected 
cluster parameters -- age, metallicity, mass, distance, reddening, and  
binary fraction -- really lead to an excellent description of the  
observed CMD, when coupled with the simulated Galactic field. 
The most noteworthy aspects in this comparison are the location and 
shape of the turn-off and subgiant branch, that   
are the features most sensitive to the cluster age. 
 
Of course, there are minor discrepancies between the observed 
and simulated data, namely: 
(i) The simulated cluster is better delineated in the CMD than  
the data. This may be ascribed to a possible underestimate of  
the photometric errors in our simulations, and to the possible  
presence of differencial reddening across the cluster (see next  
section). (ii) There is a  
deficit of simulated field stars, that can be noticed more clearly  
for $V<16$ and $(B-V)<1$. This is caused by the simplified 
way in which the thin disk is included in the Galactic model: it is 
represented by means of simple exponentially-decreasing stellar 
densities in both radial and vertical directions, and does not  
include features such as spiral arms, intervening clusters, etc., 
that are necessary to correctly describe fields at low 
galactic latitudes. Anyway, the foreground/background simulation we 
present is only meant to give us an idea of the expected location of 
field stars in the CMD.  
 
Although these shortcomings in our simulations might 
probably be eliminated with the use of slightly different 
prescriptions, they do not affect our main results, that regard the 
choice of cluster parameters. 
 
\begin{figure*} 
\centerline{\psfig{file=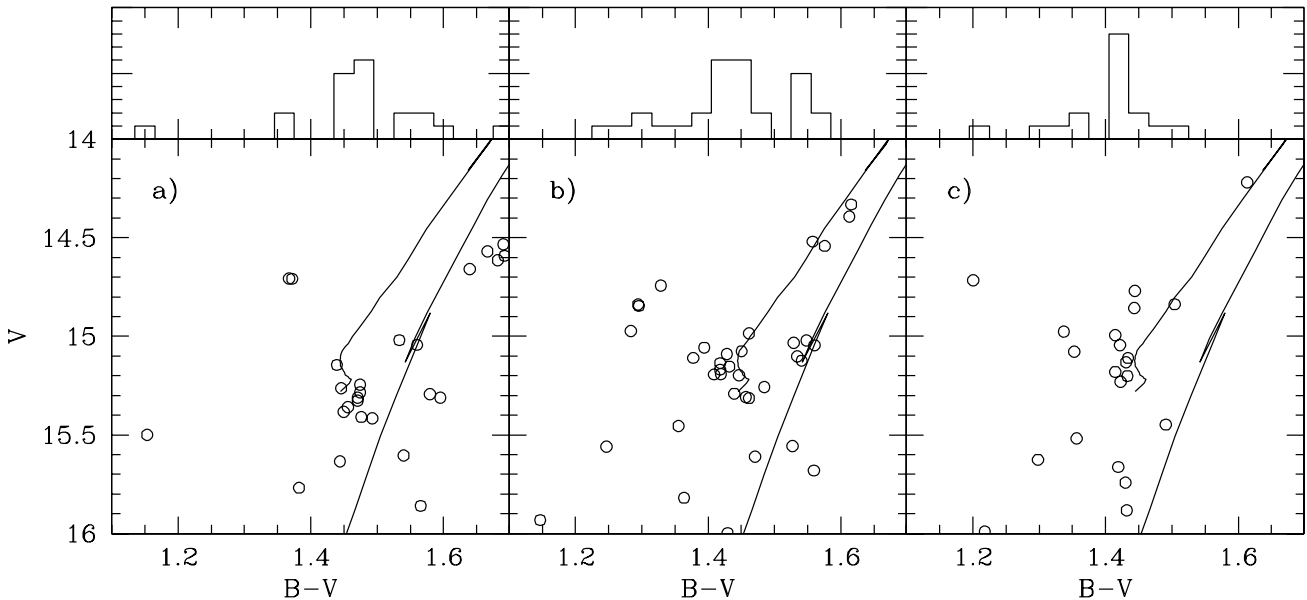,width=\textwidth}} 
\caption{ Bottom panels: The same as in Fig.~\protect\ref{fig_simulclump}, 
but now illustrating the stellar data at different intervals of  
galactic latitude, namely 
{\bf (a)} $1^{\circ}.73<b<1^{\circ}.77$ (toward NE in Fig.~\protect\ref{mappa}), 
{\bf (b)} $1^{\circ}.77<b<1^{\circ}.81$ (central part of the cluster), and 
{\bf (c)} $1^{\circ}.81<b<1^{\circ}.85$ (toward SW). 
In all cases, only stars located at $186^{\circ}.58<\ell<186^{\circ}.69$  
(a strip about $7^\prime$ wide) were plotted. As a reference to the eye,  
we plot also the same 2-Gyr old isochrone as shown in previous 
Fig.~\protect\ref{fig_isocdata}. The panels at the top illustrate, for each 
case, the colour histogram of stars with $14.7<V<15.7$. Notice the progressive  
shift of the clump to the blue as $b$ increases. 
} 
\label{fig_clump2} 
\end{figure*} 
 
 
Then, we conclude that $(m-M)_0=12.8$~mag (3630~pc), $E_{B-V}=0.60$,  
2~Gyr, $Z=0.0048$ ($[{\rm Fe/H}]=-0.60$), and $1.5\times10^4$~$M_\odot$, 
well represent the cluster parameters. All these values are  
uncertain to some extent: 
	\begin{itemize} 
	\item 
From Fig.~\ref{fig_isocdata}, we can estimate a maximum error of  
15 percent (0.3~Gyr) in the age. One should keep in mind, however,  
that the absolute age value we derived, of 2~Gyr, depends on the  
choice of evolutionary models, and specially on the prescription  
for the extent of convective cores. For the stellar masses envolved  
($M_{\rm TO}\sim1.5$~$M_\odot$ for NGC~2158),  
our models (Girardi et al. 2000a) include a moderate amount of core  
overshooting. 
	\item 
Our best fit model corresponds to $E_{B-V}=0.60$, which is compatible  
with the range $E_{B-V}=0.55\pm0.1$ indicated in 
Sect.~\ref{sec_reddening}. The uncertainty of 0.1~mag in  
$E_{B-V}$ causes an uncertainty of $\sim0.3$~mag in the  
distance modulus (15 percent in distance).  
	\item 
The metallicity cannot be better constrained from the CMD data,  
unless we have more accurate estimates of the reddening. 
	\item 
The initial mass estimate depends heavily on the choice of 
IMF, that determines the mass fraction locked into low-mass 
(unobserved) objects. The value of $1.5\times10^4$~$M_\odot$ was obtained 
with a Kroupa (2001) IMF, corrected in the lowest mass interval  
according to Chabrier (2001; details are given in Groenewegen et al. 
2001), and should be considered just as a first guess.  
At present ages, supernovae explosions and stellar mass loss  
would have reduced this mass by about 20 percent. 
	\end{itemize} 
 
\subsection{Red clump structure and differential reddening} 
 
Our cluster simulations also allows us to examine in better detail  
the observed structure of the red clump in NGC~2158. 
Figure~\ref{fig_simulclump} details the clump region in the CMD, 
for both simulations (panels a and b) and data (panel c).  
As can be readily noticed in panel (c),  
the observed clump appears as a diagonal structure, whose slope 
is roughly coincident with the reddening vector.  
In cluster simulations, however, the clump is normally seen as 
a more compact structure. This can be appreciated 
in the simulations for single stars shown in Girardi et al.\ (2000b), 
where the model clumps are found to be more elongated in the  
top-bottom direction in the CMD, and not diagonally. The same result  
is found in Figure~\ref{fig_simulclump}a, if we look only at the  
locus of single stars. 
 
 
However, part of the clump widening might be caused by the presence of 
binaries, as shown by the different symbols in our simulations 
of panels (a) and (b). It turns out that  
binaries composed of clump plus turn-off stars are located at a bluer 
colour, and are slightly brighter, if compared to the locus of single 
clump stars. Thus, binaries tend to widen 
the clump structure along a direction that roughly  
coincides with the reddening vector. But anyway, as can be readily noticed  
from Fig.~\ref{fig_simulclump}, binaries cannot account entirely for the  
clump elongation observed in NGC~2158.  
 
Instead, since NGC~2158 is located very low in the Galactic plane, 
the observed clump morphology might well be caused 
by the presence of differential reddening over the cluster. We can get  
an estimate of the expected differential reddening, starting from simple  
models of the dust distribution in the thin disk. To this aim, 
we use the Girardi et al.\ (2001) Galactic model, 
assume a local extinction value of 0.75 mag/kpc in $V$ 
(Lyng\aa\ 1982), and a diffuse dust layer of exponentially-decreasing  
density with a scale height equal to 110~pc. With these parameters, 
in correspondence of the  
NGC~2158 location we obtain a differential reddening of  
$\Delta E_{B-V}/\Delta b=-0.021$ mag/arcmin perpendicularly to the 
Galactic plane. This estimate has  
the right order of magnitude to explain the  
width of the observed clump. 
 
In order to investigate whether this kind of picture is realistic, in 
Fig.~\ref{fig_clump2} we plot the CMDs for NGC~2158, 
separated in different strips of Galactic latitude $b$. %
The 2-Gyr old isochrone, located at a fixed position in all panels,  
allows an easy visualization of how the clump gets bluer at  
increasing $b$.\footnote{A similar effect was also noticed for  
NGC~2158 main sequence stars.}. Assuming that this effect is caused by  
differential reddening perpendicularly to the Galactic plane, we get an  
estimate of $\Delta E_{B-V}/\Delta b\simeq-0.011$ mag/arcmin.  
 
We conclude that NGC~2158 presents some amount of differential  
reddening. Along the $\sim6^\prime$ diameter of the cluster, this 
effect amounts to about $\Delta E_{B-V}\sim 0.06$~mag. Since our 
previous determinations of the cluster age, distance, and reddening  
were based on the mean location of the observed stars, the correction 
of the data for differential reddening would not imply any significant  
change in the derived parameters. 
 
\section{Cluster kinematics} 
The availability of mean radial velocity and proper motion 
measurements allows us to discuss in some detail 
the kinematics of NGC~2158. Radial velocity has been measured for 
8 stars by Scott et al. (1995) and for 20 stars by Minniti (1995). 
These measurements have a comparable accuracy between 10 and 15 km/s. 
A systematic shift of about 10 km/s is likely to exist, in the sense 
that the mean radial velocity from Minniti (1995) is lower than 
that derived by Scott et al. (1995). 
Although Minniti (1995) mean radial velocity is based on better statistics,  
we shall present results based upon both the determinations. 
 
Absolute proper motions have been derived by Kharchenko et al. (1997), 
and amount to $\mu_x=+0.66\pm2.03$, $\mu_y=-3.23\pm2.16$, where 
$\mu_x=\mu_{\alpha} \cdot cos\delta$ and $\mu_y=\mu_{\delta}$. 
 
Following in details Carraro \& Chiosi (1994b) and 
Barbieri \& Gratton (2001) we derived the velocity components 
of NGC~2158 in a Galactocentric reference frame $U$, $V$ and $W$. 
The results are summarized in Table~3. 
 
\begin{table*}  
\tabcolsep 0.7truecm 
\caption {NGC~2158 basic kinematical parameters. 
The velocity components have been computed by adopting Minniti (1995, 
first row) 
Scott et al (1995, second row) radial velocity estimates.} 
\begin{tabular}{ccccccc}  
\hline 
 $(m-M)_0$ & $X$   & $Y$   & $Z$   & $U$      & $V$      & $W$ \\ 
       & kpc & kpc & kpc & km/s & km/s & km/s \\ 
\hline 
 12.80 & 12.11 &-0.42 & 0.11 & -7.99 & -56.50 & -16.88 \\ 
       &       &      &      & -21.88& -58.13 & -16.45  \\ 
\hline 
\end{tabular} 
\end{table*} 
 
By adopting the Allen \& Santill\'an (1991)  
rotationally symmetric Galaxy mass model, we integrated back 
in time NGC~2158 orbit for a duration comparable with NGC~2158 age 
(see previous Sect.~5.2),  
in order to obtain estimates of its eccentricity, 
epiciclycal ($\omega$-) and vertical ($z$-) amplitude.  
These parameters, 
together with age and metallicity, are fundamental to place the cluster 
in the right disk population. 
 
\begin{figure} 
\centerline{\psfig{file=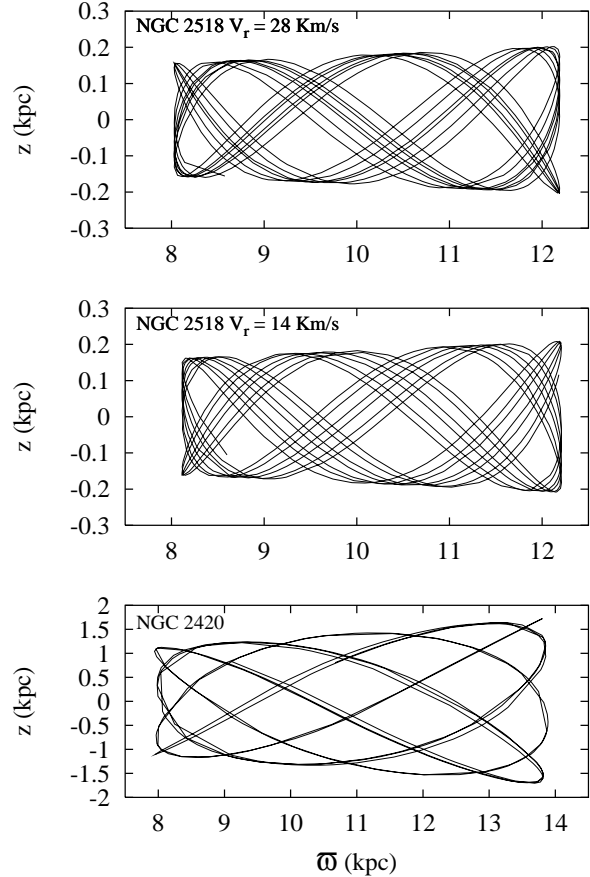,width=\columnwidth}} 
\caption{NGC~2158 orbit in  the meridional plane. In the upper 
panel we display the orbit obtained by using the radial velocity estimate 
from Scott et al. (1995), whereas the middle panel shows the  
orbit obtained by adopting the radial velocity estimate from  
Minniti (1995). The lower panel presents the orbit 
of NGC~2420.} 
\label{orbite} 
\end{figure}

The orbit integration has been performed using a modified 
version of the second-order Burlish-Stoer integrator 
originally developed by S.J. Aarseth (private communication). 
We provide orbits both for the Minniti (1995) and Scott et al. (1995) 
mean radial velocity estimates. They are shown in the upper  
and middle panels of  Fig.~\ref{orbite}. 
The parameters are basically consistent, as listed also in Table~4. 
For the sake of the discussion, in the lower panel we show  
a new orbit determination for the open cluster NGC~2420, 
which roughly shares the same age (1.8 Gyr) and metallicity  
($[{\rm Fe/H}]=-0.42$) of NGC~2158  
(Friel \& Janes 1993; Carraro et al 1998). 
The orbit of NGC~2420 was previously 
computed by Keenan \& Innanen (1974) who 
suggest that this cluster might have been disturbed in his motion 
around the Galactic center by the influence of the Magellanic Clouds, 
an hypothesis which sounds reasonable - the cluster has high eccentricity, 
large apogalacticon and stays most of the time relatively  
high above the galactic plane-  but which 
deserves a further detailed numerical investigation. 
 
Christian et al. (1985) argue about the possibility that NGC~2158 and  
NGC~2420 might share common properties and origin, since they are coeval and 
have very low metal abundances for open clusters of this age. 
It is therefore interesting to compare their orbits, also because NGC~2158 is  
even metal poorer than NGC~2420. 
With an eccentricity $e=(R_{\rm a}-R_{\rm p})/(R_{\rm a}+R_{\rm p})=0.20$ 
-- where $R_{\rm a}$ 
and $R_{\rm p}$ are the apo- and peri-galacticon, respectively -- the cluster 
reaches  a maximun distance of about 12~kpc from the Galactic Center in the 
direction of the anti-center, where it is located right now 
and where it probably formed. 
It remains relatively low in the Galactic disk, in a region  
populated by young and intermediate-age Population~I objects. 
The only difference with this population is the rather low metal content, 
less than half the solar value. 
 
Apparently, we are facing two significantly different orbits. 
NGC~2158 has an orbit more similar to normal Population I objects, 
whereas NGC~2420 possesses an eccentricity much higher than the typical  
Population I objects. Moreover, NGC~2420 is more distant than NGC~2158. 
NGC~2420 and NGC~2158 are not the only two  
cases of low metallicity intermediate-age clusters in the  
anti-center direction. 
From Carraro et al. (1998)  we have extracted 10 clusters  
with age between 1.5 and 3.5 
Gyr and low metal content ($[{\rm Fe/H}] < -0.50$).  
All these clusters presently 
lie in a Galactic sector between $\ell=135^\circ$ and $\ell=225^\circ$. 
They are: NGC~2158, NGC~2204, NGC~2420, NGC~2141, NGC~2243,  
Tombaugh~2, Berkeley~19, 20, 21, 31 and 32. 
The common properties of these clusters suggest the possibility that 
they formed from the same material. Typically, the mean metal content 
of the Galactic disk at distances between 12 and 16 kpc ranges  
between $[{\rm Fe/H}]=-0.50$ and $[{\rm Fe/H}]=-0.70$, 
according to recent estimates of the Galactic disk metallicity gradient  
(Carraro et al 1998). Such a low metal content is compatible with the low  
density, hence low star formation, probably typical of that region. 
In this respect it would be interesting to compute Galactic orbits for all  
these clusters to check whether a trend exists to have more  
eccentric orbits at increasing Galactocentric distance in this region 
of the anti-center. 
This would help to better understand the structure and evolution of the 
outer Galactic disk.

\begin{table}  
\tabcolsep 0.25truecm 
\caption {NGC~2158 orbit's basic parameters.} 
\begin{tabular}{ccccc}  
\hline 
       & $R_{\rm a}$ & $R_{\rm p}$ & $e$ & $z_{\rm max}$ \\ 
       & kpc   & kpc   &     & kpc     \\ 
\hline 
 Minniti      & 12.21 & 8.11 & 0.20 & 0.21\\ 
 Scott et al.  & 12.19 & 8.03 & 0.21 & 0.20\\ 
\hline 
\end{tabular} 
\end{table}

\section{Conclusions} 
We have presented a new CCD $UBVRI$ photometric study of the intermediate 
age open cluster NGC~2158. From the analysis of the available data we can draw 
the following conclusions: 
 
\begin{description} 
\item $\bullet$ The age of NGC~2158 is about 2 Gyr, with a 15\,\% uncertainty; 
\item $\bullet$ the reddening $E_{B-V}$ turns out to be $0.55\pm0.10$ 
and we find evidence of differential reddening (of about 0.06~mag)  
across the cluster; 
\item $\bullet$ we place the cluster at about 3.6 kpc from the Sun toward 
the anti-center direction; 
\item $\bullet$ combining together NGC~2158 age, metallicity and kinematics, 
we suggest that it is a genuine member of the old thin disk population. 
\end{description}

\section*{Acknowledgements} 
We are very grateful to Chiara Miotto for carefully reading 
this manuscript, to Mauro Barbieri for NGC~2158 orbit integration, 
to Martin Groenewegen for the latest calibration of the 
Galactic model, and to Luciano Traverso, who secured the observations
of January 7. We acknowledge
also the referee, dr. G. Gilmore, for his useful suggestions.
This study has been financed by the Italian Ministry of 
University, Scientific Research and Technology (MURST) and the Italian 
Space Agency (ASI), and made use of Simbad and WEBDA databases.

\end{document}